\documentclass[prl,aps,twocolumn,floatfix,superscriptaddress,showpacs]{revtex4-1}
\usepackage{graphicx,graphics,psfrag,amsmath,amsfonts,calc,epsfig,color}

\begin{document}

\title{Effects of spin-orbit coupling on the Berezinskii-Kosterlitz-Thouless transition and the vortex-antivortex structure in two-dimensional Fermi gases}
	
\author{Jeroen P.A. Devreese}
\address{School of Physics, Georgia Institute of Technology, 
Atlanta, 30332, USA}
\address{TQC, Universiteit Antwerpen, B-2610 Antwerpen, Belgium}

\author{Jacques Tempere}
\address{TQC, Universiteit Antwerpen, B-2610 Antwerpen, Belgium}
\affiliation{Lyman Laboratory of Physics, Harvard University, Cambridge, Massachusetts 02138, USA}

\author{Carlos A.R. S\'a de Melo}
\address{School of Physics, Georgia Institute of Technology, 
Atlanta, 30332, USA}

\date{\today}

%-----------------------------------------------------------------%
\begin{abstract}
We investigate the Berezinskii-Kosterlitz-Thouless (BKT) transition in a two-dimensional (2D) Fermi gas with spin-orbit coupling (SOC), as a function of the two-body binding energy and a perpendicular Zeeman field. By including a generic form of the SOC, as a function of Rashba and Dresselhaus terms, we study the evolution between the experimentally relevant equal Rashba-Dresselhaus (ERD) case and the Rashba-only (RO) case. We show that in the ERD case, at fixed non-zero Zeeman field, the BKT transition temperature $T_{BKT}$ is increased by the effect of SOC for all values of the binding energy. We also find a significant increase in the value of the Clogston limit compared to the case without SOC. Furthermore, we demonstrate that the superfluid density tensor becomes anisotropic (except in the RO case), leading to an anisotropic phase-fluctuation action that describes elliptic vortices and antivortices, which become circular in the RO limit. This deformation constitutes an important experimental signature for superfluidity in a 2D Fermi gas with ERD SOC. Finally, we show that the anisotropic sound velocities exhibit anomalies at low temperatures, in the vicinity of quantum phase transitions between topologically distinct uniform superfluid phases.
\end{abstract}
%-----------------------------------------------------------------%

\pacs{%Degenerate Fermi gas%
03.75.Ss,
%Fermi gas/Electron gas quantum statistical mechanics% 
05.30.Fk,
%Hydrodynamic aspects of superfluidity% 
47.37.+q,
%Vortex phases (superconductivity)% 
74.25.Uv,
%Kosterlitz-Thouless transition in magnetic systems% 
75.30.Kz}

\maketitle

\begin{figure}[t]
\centerline{{\includegraphics[keepaspectratio=true,width=86mm]{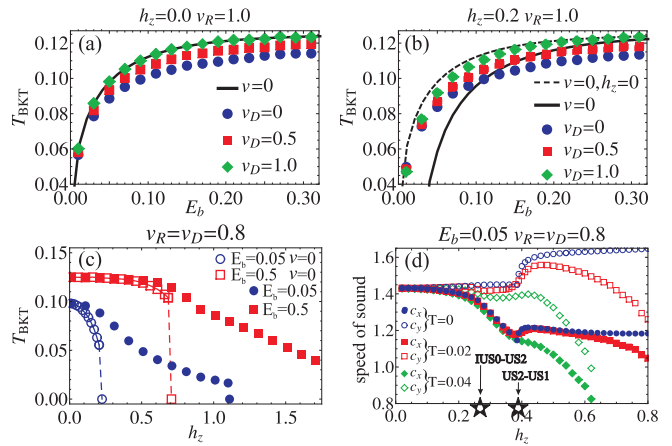}}}
\caption{(Color online) Berezinskii-Kosterlitz-Thouless temperature $T_{BKT}/E_F$ as a function of the two-body binding energy $E_b/E_F$, at fixed values of the Zeeman field $h_z/E_F$: (a) $h_z/E_F=0$ and (b) $h_z/E_F=0.2$. The Rashba coupling strength is held fixed at $v_R/\tilde{v}_F=1$ (with $\tilde{v}_F=v_F/2)$, while the Dresselhaus coupling strength $v_D/\tilde{v}_F$ is varied. The reference case without SOC ($v=0$) is also shown. (c) $T_{BKT}/E_F$ as a function of $h_z/E_F$ for ERD SOC ($v_R/\tilde{v}_F=v_D/\tilde{v}_F=0.8$) and the reference case without SOC ($v=0$) at fixed $E_b$. (d) The sound velocities $c_x/\tilde{v}_F$ and $c_y/\tilde{v}_F$ for ERD SOC ($v_R/\tilde{v}_F=v_D/\tilde{v}_F=0.8$) as a function of $h_z$, at fixed $E_b$ and varying $T$. The quantum phase transitions at $T=0$ between the indirect gapped state (i-US-0) and the gapped state with four (US-2) and two (US-1) nodes are indicated.}
\label{Fig1}
\end{figure}

\begin{figure}[t]
\centerline{{\includegraphics[keepaspectratio=true,width=86mm]{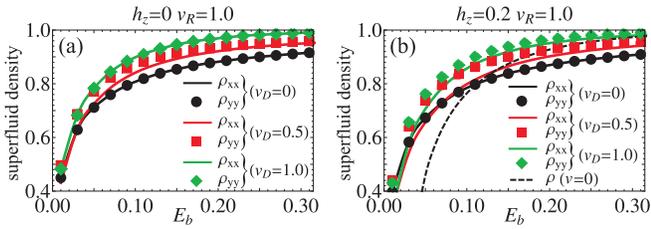}}}
\caption{(Color online) The components of the anisotropic superfluid density tensor, $\rho_{xx}/E_F$ and $\rho_{yy}/E_F$ as a function of the binding energy $E_b/E_F$, for fixed Rashba coupling strength $v_R/\tilde{v}_F=1$ and varying Dresselhaus coupling strength $v_D/\tilde{v}_F$: (a) $h_z/E_F=0$ and (b) $h_z/E_F=0.2$.}
\label{Fig2}
\end{figure}

\begin{figure}[t]
\centerline{{\includegraphics[keepaspectratio=true,width=86mm]{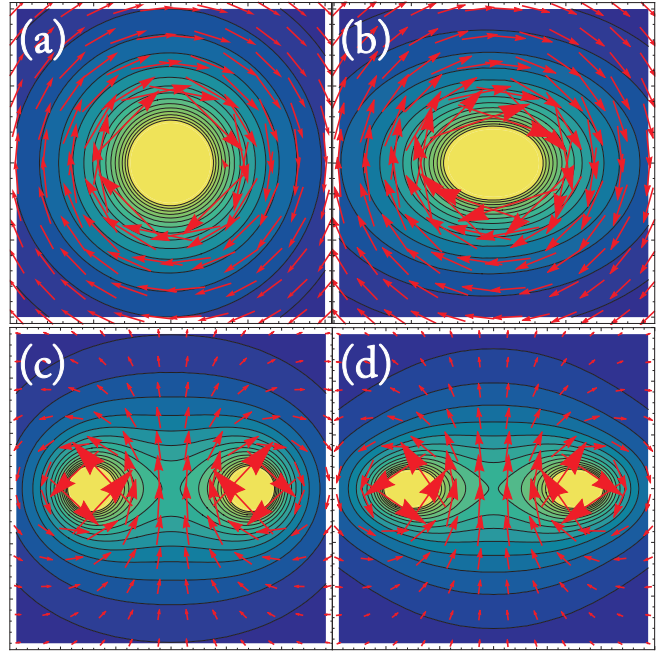}}}
\caption{\label{Fig3}(Color online) Single antivortex and vortex-antivortex structure for Rashba-only SOC [(a) and (c)], and equal Rashba-Dresselhaus SOC [(b) and (d)]. In the former case, the antivortex exhibits circular symmetry around its core, whereas in the latter case the antivortex is elongated along one axis and contracted along the other. The parameters used are $h_z/E_F=0.2$, $E_b/E_F=0.01$, $T\approx T_{BKT}$, with $v_R/\tilde{v}_F=1$ and $v_D/\tilde{v}_F=0$ in (a) and (c), and $v_R/\tilde{v}_F=v_D/\tilde{v}_F=1$ in (b) and (d).}
\end{figure}

%Introduction
%-----------------------------------------------------------------%
In the last few years, ultracold gases have been used as quantum simulators to study many-body systems. Presently, several quantities, such as the interaction strength, population-imbalance and dimensionality are controllable parameters in experiments with ultracold atoms. Recently, the ability to engineer artificial gauge fields has been added to this repertoire \cite{Spielman artificial magnetic fields Bosons (2009),Review Dalibard (2011)}. One example of this is the creation of spin-orbit coupling (SOC) in neutral atoms. This capability not only has the potential to shed more light on condensed matter systems such as topological insulators \cite{Kane RMP (2010)}, but also to create new states of matter that have no analogue in other fields of physics, such as synthetic spin-$1/2$ bosons \cite{Leggett (2003)}. In ultracold atomic gases, the only SOC that has been experimentally created so far is the equal Rashba-Dresselhaus (ERD) case, which was first achieved in systems of trapped interacting bosons \cite{Spielman SOC Bosons (2011)} and subsequently in systems of trapped non-interacting fermions \cite{Zhang SOC Fermions (2012),Zwierlein SOC Fermions (2012)}. Recently, the interacting spin-orbit coupled Fermi gas near a Feshbach resonance has also been realized \cite{Spielman SOC Fermions with interaction (2013)}.\newline\indent On the theoretical side, most investigations of SOC in ultracold fermionic gases were focused on three-dimensional (3D) systems with either Rashba-only (RO) coupling \cite{Rashbons,Iskin (2011),Zhang (2011),Pu (2011),Zhai (2011)} or ERD coupling \cite{Sa de Melo 3D (1) (2012),Sa de Melo 3D (2) (2012)}. The RO case has a strong connection to the condensed matter literature \cite{Cond.-mat.}, in which the ERD case has no counterpart. In the field of ultracold atoms, the two-dimensional (2D) RO case has also received attention \cite{Chuanwei Zhang (Jan 2012),Zhidong Zhang (2012)}, in part because of its relation to topological phase transitions \cite{Yi (2011)} and the emergence of Majorana zero-energy modes \cite{Chuanwei Zhang (Sep 2012)}. Additionally, the ERD case has been studied at the mean-field level in the zero temperature limit \cite{Han Sa de Melo arxiv}. However, it is well known that in 2D the finite temperature superfluid transition arises due to the Berezinskii-Kosterlitz-Thouless (BKT) mechanism of vortex-antivortex unbinding \cite{Berezinskii,Kosterlitz}, where phase fluctuations play a fundamental role. Hence, in 2D, it is essential to go beyond the mean-field approximation and include phase fluctuations \cite{Botelho_SadeMelo(2006),Klimin Tempere (2009),X-G Huang (2012)} when dealing with finite temperatures.
\newline \indent In this manuscript, we study the BKT transition in a 2D Fermi gas in the presence of SOC. We use a generic form of SOC, which allows us to investigate the evolution between the experimentally relevant ERD case and the isotropic RO case. We determine the BKT transition temperature $T_{BKT}$ as a function of the two-body binding energy and a perpendicular Zeeman field and we discuss the vortex-antivortex structure of the system. The main results of our work can be summarized as follows: (1) in the ERD case, at fixed non-zero Zeeman field, $T_{BKT}$ is increased by the presence of SOC, for all values of the binding energy. (2) SOC leads to an anisotropic superfluid density tensor (except in the RO case). The resulting anisotropic fluctuation action describes elliptic vortices and antivortices, which become circular in the RO limit. (3) The anisotropic sound velocities are sensitive to the presence of quantum phase transitions, where the uniform superfluid (US) state changes its nodal structure from fully gapped (US-0) to gapless with four (US-2) or two nodes (US-1). \newline
%Main calculation
%-----------------------------------------------------------------%
\indent \textit{Hamiltonian.}\textemdash The starting point of our analysis is the Hamiltonian density: $\mathcal{H}(r)=\mathcal{H}_{0}(r) + \mathcal{H}_{SOC}(r) + \mathcal{H}_{int}(r)$, where $r=(\textbf{r},\tau)$ is a three-vector representing position $\textbf{r}$ and imaginary time $\tau$. The single-particle component is given by $\mathcal{H}_{0}(r)=\sum_{s,s'}\bar{\psi}_{r,s}\left( \hat{K}_s\delta_{s,s'}-h_{z}\sigma_{z,ss'} \right)\psi_{r,s'}$, where $\hat{K}_s=-\nabla_\textbf{r}^2-\mu_{s}$ is the kinetic energy in reference to the chemical potential $\mu_{s}$ of the fermions in spin state $s=(\uparrow,\downarrow)$. Furthermore, $h_{z}$ is the Zeeman field along the $z$-axis perpendicular to the $(x,y)$-plane, $\sigma_i$ indicates the $i^{th}$ Pauli matrix and $\bar{\psi}_{r,s}$ and $\psi_{r,s}$ are fermionic fields. The second term
\begin{align}
\mathcal{H}_{SOC}(r)=-2\sum_{s,s'}\bar{\psi}_{r,s} ( \alpha\hspace{0.5mm} \hat{k}_{x}\sigma_{y,ss'}-\gamma \hspace{0.5mm} \hat{k}_{y}\sigma_{x,ss'}) \psi_{r,s'}
\end{align}
is the SOC part, where $\hat{k}_{\textbf{r}_i}=-i(\partial/\partial {\textbf{r}_i})$ and with $\alpha=(v_{R}+v_{D})/2$ and $\gamma=(v_{R}-v_{D})/2$ being the sum and difference of the Rashba and Dresselhaus coupling strength $v_{R}$ and $v_{D}$, respectively. We have written the SOC part in a generic form, which allows us to study the evolution from the ERD case ($v_R=v_D$) to the RO case ($v_D=0$). Finally, the third term $\mathcal{H}_{int}(r)=g\bar{\psi}_{r,\uparrow}\bar{\psi}_{r,\downarrow}\psi_{r,\downarrow}\psi_{r,\uparrow}$, is a local interaction Hamiltonian, where $g$ is the strength of the contact interaction. For the remainder of this work we will use the units $\hbar=2m=E_F=1$, with $m$ and $E_F$ being the atomic mass and the Fermi energy, respectively. \newline
\indent \textit{Functional integral derivation.}\textemdash The main goal of our calculation is to derive the effective action of the system, from which quantities such as the superfluid density, the compressibility and the vortex-antivortex structure can be readily obtained. To arrive at this action, we perform a functional integration over the fermionic fields, leading to the partition function: $\mathcal{Z}=\int \mathcal{D}\bar{\psi}\mathcal{D}\psi \exp[-S(\bar{\psi},\psi)]$. The action $S(\bar{\psi},\psi)$ of the system is linked to the Hamiltonian density via a Legendre transformation: $S(\bar{\psi},\psi)=\int dr \left[ \sum_{s}\bar{\psi}_{r,s} \frac{\partial}{\partial\tau} \psi_{r,s} + \mathcal{H}(r) \right].$ To decouple the fourth order interaction term in $\mathcal{H}_{int}$ into second order terms, we use the standard Hubbard-Stratonovich transformation. This introduces a functional integral over the complex pair fields $\bar{\Delta}_r$ and $\Delta_r$. Subsequently, we introduce the phase of the order parameter into the system by re-writing: $\Delta_r=|\Delta_r|e^{i\theta_r}$, where $|\Delta_r|$ is the amplitude and $\theta_r$ is the phase. Furthermore, to make explicit the dependence of the action on the phase $\theta_r$, we perform the gauge-transformation: $\psi_{r,s}\rightarrow\psi_{r,s}e^{i\theta_r/2}$. \newline \indent Integration over the fermionic fields $\bar{\psi}_{r,s}$ and $\psi_{r,s}$ in the partition function $\mathcal{Z}$ leads to an effective action that is a function of $|\Delta_r|$ and $\theta_r$. However, since phase fluctuations provide the dominant contributions to the physics in 2D, we can take the amplitude of the order parameter to be uniform in space and imaginary time $(|\Delta_r|=|\Delta|)$. Using this procedure, we arrive at the partition function $\mathcal{Z}=\int\mathcal{D}\theta_r\exp[-S(\theta_r)]$, where the action is given by
\begin{align}\label{Action after functional integration over Delta and psi}
S=&-\frac{1}{2} {\rm Tr} \left\{\ln\left[\beta\begin{pmatrix}
\mathbb{A}_+&&\mathbb{D}_+\\
\mathbb{D}_-&&\mathbb{A}_-^*
\end{pmatrix}\right]\right\}-\frac{\beta L^2 |\Delta|^2}{g}\\
&+\frac{\beta}{2}\sum_{k,s}(-i\omega_n+\textbf{k}^2-\mu_s)+\frac{1}{8 L^2}\int dr\sum_k[\nabla_\textbf{r}(\theta_r)]^2\nonumber.
\end{align}
In this expression, $\beta$ is the inverse temperature, $L^2$ is the area of the 2D system and $k=(\textbf{k},\omega_n)$ is a three-vector representing the fermionic wave vector $\textbf{k}$ and the fermionic Matsubara frequency $\omega_n=(2n+1)\pi/\beta$. The matrix in (\ref{Action after functional integration over Delta and psi}) is a $4\times 4$ matrix, written as a function of the $2\times 2$ matrices
\begin{align}\label{submatrix A}
\mathbb{A}_\pm=
\begin{pmatrix}
\mp i\omega_n\pm\xi_\textbf{k}^\theta\mp\widetilde{\zeta}-\zeta_\textbf{k}^\theta&&-h_{\bot}(\textbf{k})\mp h_{\bot}^{\theta}\\
-h^*_{\bot}(\textbf{k})\mp h_{\bot}^{*\theta}&&\mp i\omega_n\pm\xi_\textbf{k}^\theta\pm\widetilde{\zeta}-\zeta_\textbf{k}^\theta
\end{pmatrix}
\end{align}
and $\mathbb{D}_{\pm}=\pm i\sigma_y|\Delta|$. The kinetic terms in (\ref{submatrix A}) have been divided into phase-independent and phase-dependent contributions, where we defined $\xi_\textbf{k}^\theta=\xi_\textbf{k}+\xi^\theta$. The phase-independent terms are $\xi_\textbf{k}=\textbf{k}^2-\mu$ with $\mu=(\mu_\uparrow+\mu_\downarrow)/2$ and $\widetilde{\zeta}=\zeta+h_z$ with $\zeta=(\mu_\uparrow-\mu_\downarrow)/2$. The phase-dependent terms are $\xi^\theta=\frac{i}{2}\frac{\partial\theta_r}{\partial\tau}+\frac{1}{4}[\nabla_\textbf{r}(\theta_r)]^2$ and $\zeta_\textbf{k}^\theta=-\nabla_\textbf{r}(\theta_r)\cdot\textbf{k}$. The spin-flip terms also contain a phase-independent contribution corresponding to the SOC field $h_{\bot}(\textbf{k})=-2\gamma\hspace{0.5mm}k_y-2i\alpha\hspace{0.5mm}k_x$ and a phase-dependent contribution $h_{\bot}^\theta=-\gamma\frac{\partial\theta_r}{\partial y}-i\alpha\frac{\partial\theta_r}{\partial x}$.
\newline
\indent \textit{Quadratic expansion of the action.}\textemdash The final step in obtaining the effective action is to perform an expansion of expression (\ref{Action after functional integration over Delta and psi}) to quadratic order in the phase $\theta_r$. Denoting the $4\times 4$ matrix appearing in the action (\ref{Action after functional integration over Delta and psi}) by $\mathbb{M}_k(\theta,\partial\theta)$, we can write ${\rm Tr}\{\ln[\beta\mathbb{M}_k(\theta,\partial\theta)]\}={\rm Tr}\{\ln[\beta\mathbb{M}_k(0,0)]\}+{\rm Tr}\{\ln[\mathbb{I}+\mathbb{M}^{-1}_k(0,0)\mathbb{F}_k(\theta,\partial\theta)]\}$, where we have defined $\mathbb{F}_k(\theta,\partial\theta)=\mathbb{M}_k(\theta,\partial\theta)-\mathbb{M}_k(0,0)$ to be the phase-fluctuation part. The first term in the expansion leads to the saddle-point part of the action $S_{sp}=-\frac{1}{2}{\rm Tr}\{\ln[\beta\mathbb{M}_k(0,0)]\}+\frac{\beta}{2}\sum_{k,s}(-i\omega_n+\textbf{k}^2-\mu_s)-\frac{\beta L^2 |\Delta|^2}{g}$, which results in the saddle-point thermodynamic potential
\begin{align}\label{saddle-point thermodynamic potential}
\Omega_{sp}=\sum_\textbf{k}\left(\frac{-1}{2\beta}\sum_{i=\pm}\ln[2+2\cosh(\beta E_i)]+\xi_\textbf{k}\right)-\frac{L^2|\Delta|^2}{g}.
\end{align}
In this expression we used the eigenvalues of the matrix $\mathbb{M}_k(0,0)$, which are given by $E_\pm=\sqrt{\epsilon_{\textbf{k}}^2+\vartheta^2\pm2\sqrt{\epsilon_{\textbf{k}}^2\vartheta^2-|\Delta|^2|h_{\bot}(\textbf{k})|^2}}$ where $\epsilon_\textbf{k}^2=\xi_\textbf{k}^2+|\Delta|^2$ and $\vartheta^2=\widetilde{\zeta}^2+|h_{\bot}(\textbf{k})|^2$. Furthermore, in 2D, the coupling strength $g$ can be eliminated in favor of the two-body binding energy $E_b$ through the relation: $\frac{1}{g}=-\int\frac{d\textbf{k}}{(2\pi)^2}\frac{1}{2k^2+E_b}$.\newline
The second term in the expansion leads to the phase-fluctuation action: $S_{fl}=-\frac{1}{2}{\rm Tr}\{\ln[\mathbb{I}+\mathbb{M}^{-1}_k(0,0)\mathbb{F}_k(\theta,\partial\theta)]\}+\frac{1}{8 L^2}\int dr\sum_k[\nabla_\textbf{r}(\theta_r)]^2$, which becomes
\begin{align}\label{fluctuation action}
S_{fl}=\frac{1}{2}\int dr \left(\mathcal{A}\left(\frac{\partial\theta_r}{\partial\tau}\right)^2+\sum_{\nu=\{x,y\}}\rho_{\nu\nu}\left( \frac{\partial\theta_r}{\partial \nu} \right)^2 \right),
\end{align}
after explicit expansion in $\nabla_\textbf{r}(\theta_r)$. The exact expressions for the compressibility $\mathcal{A}$ and the superfluid density tensor components $\rho_{xx}$ and $\rho_{yy}$ are given in the supplemental material \cite{supplementary material}. Here, we note the symmetry relation $\rho_{xx}(v_R,v_D)=\rho_{yy}(v_R,-v_D)$ together with the fact that $\rho_{xx}\neq\rho_{yy}$, provided that the SOC is anisotropic. The difference between $\rho_{xx}$ and $\rho_{yy}$ is a direct consequence of the anisotropy in the higher-angular-momentum pairing of the SOC-induced triplet component of the order parameter.\newline
%Results and discussion
%-----------------------------------------------------------------%
\indent \textit{The phase diagram and sound velocities.}\textemdash For the remainder of this work, we particularize to the population balanced case with $\mu_\uparrow=\mu_\downarrow=\mu$. In order to determine the finite temperature phase diagram, we need to find the amplitude $|\Delta|$ of the order parameter, the chemical potential $\mu$ and the Berezinskii-Kosterlitz-Thouless transition temperature $T_{BKT}$. This leads to a set of three equations that need to be solved self-consistently: (1) the order parameter equation, determined by the condition $\partial\Omega_{sp}/\partial\Delta=0$, (2) the number equation $-\partial\Omega_{sp}/\partial\mu=n$, and (3) the generalized Kosterlitz-Thouless condition $T_{BKT}=\frac{\pi}{2}\rho_s(T_{BKT})$, where $\rho_s=\sqrt{\rho_{xx}\rho_{yy}}$. The first two equations define the mean-field (saddle-point) ``transition'' temperature $T_{MF}$, at which the system undergoes a ``transition'' between the normal state $|\Delta|=0$ and the paired state $|\Delta|\neq 0$. However, the transition to a true superfluid state occurs at $T_{BKT}<T_{MF}$, which is greatly affected by phase fluctuations. Determining this critical temperature requires the simultaneous solution of all three aforementioned equations.\newline
\indent Figures \ref{Fig1}(a) and (b) show the self-consistency solution for $T_{BKT}$ as a function of the binding energy $E_b$, for fixed Zeeman field $h_z$. The Rashba coupling strength $v_R$ is held fixed in a given figure at $v_R/\tilde{v}_F=1$ (with $\tilde{v}_F=v_F/2$ and $v_F$ the Fermi velocity), while the Dresselhaus coupling strength $v_D/\tilde{v}_F$ is varied. This allows us to study the evolution between the currently experimentally relevant ERD case and the RO case. In Fig. \ref{Fig1}(a), where $h_z=0$, the highest $T_{BKT}$ is always achieved in the case without SOC ($v=0$, black solid line). This is because SOC tends to make zero center-of-mass momentum pairing more difficult by introducing orbital frustration. In the ERD case, however, this orbital frustration can be removed by a gauge transformation, hence $T_{BKT}$ (green diamonds) does not decrease. In Fig. \ref{Fig1}(b), where $h_z\neq 0$, the presence of SOC offsets the effect of the Zeeman field, by introducing a triplet pairing component. In the ERD case (green diamonds), $T_{BKT}$ increases for all values of $E_b$, compared to the case without SOC (black solid line). In the `hybrid' case (red squares) and the RO case (blue circles), an increase of $T_{BKT}$ occurs for small values of $E_b$. For large values of $E_b$, however, the orbital frustration effect becomes dominant and $T_{BKT}$ decreases compared to the case without SOC.
\newline \indent In Fig. \ref{Fig1}(c), $T_{BKT}$ is plotted as a function of the Zeeman field $h_z$ for the ERD case $v_R/\tilde{v}_F=v_D/\tilde{v}_F=0.8$, together with the reference case without SOC $(v=0)$, for two values of the binding energy: $E_b/E_F=0.05$ (blue empty and solid circles) and $E_b/E_F=0.5$ (red empty and solid squares). This figure shows that ERD SOC stabilizes the superfluid state against the effect of a Zeeman field. As a result, the Clogston limit, at which $T_{BKT}$ jumps discontinuously to zero, lies at a significantly higher value in the case with SOC ($h_z^{(c)}/E_F\approx 1.1$ for $E_b/E_F=0.05$) compared to when no SOC is present ($h_z^{(c)}/E_F\approx 0.21$ for $E_b/E_F=0.05$). This larger generalized Clogston limit occurs in the phase locked case $(\theta_{r,\uparrow}=\theta_{r,\downarrow})$, which becomes orbitally frustrated, even though there is an induced triplet component of the order parameter. 
\newline \indent In Fig. \ref{Fig1}(d), we show the speed of sound along the $x$ and $y$ directions as a function of $h_z$, for the ERD case at low temperatures. Notice that for $E_b/E_F\ll 1$ and $h_z=0$, the sound velocities $c_x/\tilde{v}_F=c_y/\tilde{v}_F=\sqrt{2}$ reduce to the standard value in the BCS regime given by $c_x=c_y=v_F/\sqrt{2}$. In Fig. \ref{Fig1}(d), we have indicated the quantum phase transitions between the topologically distinct phases at $T=0$. In the ERD case, the uniform superfluid (US) phases can be classified according to the nodal structure of the quasiparticle energies $E_\pm$ (see (\ref{saddle-point thermodynamic potential})), leading to four cases \cite{Han Sa de Melo arxiv}. The first two cases correspond to a uniform superfluid phase with no nodes, which can have an indirect gap (i-US-0) at non-zero momentum, or a direct gap (d-US-0) at zero momentum. The last two cases correspond to uniform superfluid phases with two nodes (US-1) or four nodes (US-2). Fig. \ref{Fig1}(d) shows that the sound velocity at low temperature is sensitive to the quantum phase transition (QPT) between the US-2 and US-1 phases. The reason for this sensitivity is that at this transition two nodal Dirac quasiparticles with opposite topological charges annihilate at zero momentum, i.e., in the long-wave length limit. However, the sound velocities are much less sensitive to the QPT between the i-US-0 and US-2 phases, because in that case the transition occurs at finite quasiparticle momenta. When approached from the i-US-0 side, this transition can be viewed as the softening of the quasiparticle excitation spectrum of the gapped i-US-0 phase at two finite momenta near $\pm k_F$. When approached from the US-2 side, the transition can be understood as the annihilation of two Dirac quasiparticles
with opposite topological charges at non-zero momentum.
\newline
\indent \textit{Anisotropy and vortex-antivortex structure.}\textemdash An important effect of SOC is that the superfluid density tensor can become anisotropic, as is shown in expression (\ref{fluctuation action}). In Fig. \ref{Fig2}, we show the components $\rho_{xx}$ and $\rho_{yy}$ of the superfluid density, as a function of $E_b$, for fixed Rashba coupling strength $(v_R/\tilde{v}_F=1)$ and varying Dresselhaus coupling strength $v_D/\tilde{v}_F$. In Fig. \ref{Fig2}(a), $\rho_{xx}=\rho_{yy}$ in two situations: (1) in the ERD case (green diamonds and green solid line), where the effect of SOC can be gauged away because $h_z=0$, and (2) in the RO case (black circles and black solid line), which is isotropic. In Fig. \ref{Fig2}(b), we show that $\rho_{yy}>\rho_{xx}$ when $h_z\neq0$, provided that the SOC is anisotropic.
\newline
\indent The SOC-induced anisotropy of the superfluid density
has important effects on the vortex structure of the superfluid state. In the case without SOC or in the RO case, $\rho_{xx}=\rho_{yy}$ and thus the vortex (and antivortex) solutions exhibit circular symmetry, as shown in Fig. \ref{Fig3}(a). However, in the presence of anisotropic SOC, such as the ERD case, the superfluid density tensor is also anisotropic, and leads to elliptic rather than circular vortices. This effect is demonstrated in Fig. \ref{Fig3}(b). The general solutions used in Figs. \ref{Fig3}(a) and (b) are of the form $\theta_{V}(x,y)=\pm\arctan(\tilde{\rho}^2y/x)$ with $\tilde{\rho}=(\rho_{xx}/\rho_{yy})^{1/4}$ for a vortex (antivortex) located at coordinates $(x=0,y=0)$. 
The general solution for the vortex-antivortex (VA) pair is $\theta_{VA}(x,y)=\arctan[2\tilde{a}\tilde{y}/(\tilde{a}^2-\tilde{x}^2-\tilde{y}^2)]$, where $\tilde{x}=x/\tilde{\rho}$, $\tilde{y}=y\tilde{\rho}$ and $\tilde{a}=a/\tilde{\rho}$. The parameter $a$ indicates half the VA pair size: the location of the vortex/antivortex is at $(x=\pm a,y=0)$. Plots of these solutions are shown in Fig. \ref{Fig3}(c) and (d) for the RO case and for the ERD case, respectively. The parameters used in Fig. \ref{Fig3} are $E_B/E_F=0.01$ and $h_z/E_F=0.2$. We chose these parameters to enhance visualization, as the ratio of $\rho_{yy}/\rho_{xx}$ is larger for smaller binding energy, as can be seen from Fig. \ref{Fig2}. The emergence of elliptic vortices and the structure of the VA pairs in a 2D Fermi superfluid constitute important signatures for the experimentally relevant ERD case. These signatures could be detected during a time-of-flight expansion of the trapped system, or via Bragg spectroscopy, which is also sensitive to the direction of rotation of the supercurrents.\newline
%Conclusion
%-----------------------------------------------------------------%
\indent \textit{Conclusion.}\textemdash We have studied the effects of spin-orbit coupling (SOC) with an arbitrary mixture of Rashba and Dresselhaus terms on the Berezinskii-Kosterlitz-Thouless (BKT) transition in a 2D Fermi gas. We have found that in the equal Rashba-Dresselhaus (ERD) case, at fixed non-zero Zeeman field, the BKT transition temperature $T_{BKT}$ increases for all values of the binding energy, due to the emergence of an SOC-induced triplet component of the
order parameter. However, $T_{BKT}$ never becomes larger that the case of vanishing Zeeman and SOC fields, because of residual orbital effects. In addition, we found a significant increase in the value of the Clogston limit, compared to the case without SOC. Furthermore, we have shown that at non-zero Zeeman field, the superfluid density tensor becomes anisotropic due to the presence of SOC (except in the Rashba-only case). This anisotropy leads to vortices which exhibit elliptical symmetry in the experimentally relevant ERD case. This deformation of vortices and antivortices constitutes an important experimental signature for superfluidity in a 2D Fermi gas with anisotropic SOC. Finally, we have shown that the anisotropic sound velocities exhibit anomalies near the quantum phase transition between the uniform superfluid (US) phase with four nodes (US-2) and two nodes (US-1).
%Acknowledgement
%-----------------------------------------------------------------%
\acknowledgements{One of us (JPAD) wishes to thank E. Vermeyen, G. Lombardi and D. Sels for interesting and stimulating discussions. JPAD gratefully acknowledges a Post-doctoral fellowship of the Research Foundation-Flanders (FWO-Vlaanderen). This project was supported by projects G.0370.09N, G.0180.09N, G.0119.12N, G.0122.12N, WOG (WO.035.04N) (JT) and ARO (W911NF-09-1-0220) (CARS).}

%Bibliography
%-----------------------------------------------------------------%

\end{document}